# A Model-Driven Engineering Approach to AI-Powered Healthcare Platforms

MIRA, RAHEEM

Faculty of Computers & Artificial Intelligence, Cairo University, Cairo, Egypt.

Scientific Academy for Service Technology e.V. (ServTech), Potsdam, Germany

AMAL, ELGAMMAL

Egypt University of Informatics, New Administrative Capital, Cairo, Egypt.

Faculty of Computers & Artificial Intelligence, Cairo University, Cairo, Egypt

Scientific Academy for Service Technology e.V. (ServTech), Potsdam, Germany

MICHAEL, PAPAZOGLOU

Scientific Academy for Service Technology e.V. (ServTech), Potsdam, Germany

University of New South Wales, Sydney, Australia

University of Macquarie, Sydney, Australia

BERND, KRÄMER

Scientific Academy for Service Technology e.V. (ServTech), Potsdam, Germany

NEAMAT, EL-TAZI

Faculty of Computers & Artificial Intelligence, Cairo University, Cairo, Egypt

Artificial intelligence (AI) has the potential to transform healthcare by supporting more accurate diagnoses and personalized treatments. However, its adoption in practice re-mains constrained by fragmented data sources, strict privacy rules, and the technical complexity of building reliable clinical systems. To address these challenges, we in-troduce a model-driven engineering (MDE) framework designed specifically for healthcare AI. The framework relies on formal metamodels, domain-specific languages (DSLs), and auto-mated transformations to move from high-level specifications to run-ning software. At its core is the Medical Interoperability Language (MILA), a graphical DSL that enables clinicians and data scientists to define que-ries and machine learning pipelines using shared ontologies. When combined with a federated learning ar-chitec-ture, MILA allows institutions to collaborate without exchanging raw patient data, ensuring semantic consistency across sites while preserving privacy. We evaluate this approach in a multi-center cancer immu-notherapy study. The generated pipelines de-livered strong predictive performance, with support vector ma-chines achieving up to 98.5% and 98.3% accuracy in key tasks, while substantially reducing manual coding ef-fort. These findings suggest that MDE principles—metamodeling, semantic integration, and automated code generation—can provide a practical path toward interoperable, reproducible, and trustworthy digital health platforms.

## 1 INTRODUCTION

The intersection of digital health and artificial intelligence (AI) is reshaping modern healthcare. Tools ranging from diagnostic decision support to personalized treatment planning, population-level analytics, and remote patient monitoring are now within reach. At the same time, putting these technologies into practice is far from straightforward. Clinical data are often fragmented across institutions, stored in heterogeneous formats, and governed by strict privacy rules such as GDPR and HIPAA. These barriers limit interoperability, complicate the development of AI models, and make traceability essential for both safety and regulatory compliance [1], [2].

Model-driven engineering (MDE) provides one possible way forward. Instead of building systems entirely through code, MDE raises the level of abstraction: clinical workflows and domain concepts are expressed as formal models that can be automatically turned into running software. With the help of domain-specific languages (DSLs), clinicians and data scientists can describe analytical tasks in familiar terms, while the underlying MDE process ensures modularity, traceability, and reusability. These qualities are particularly important for building healthcare AI systems that must be safe, adaptable, and auditable [3], [4].

**Research Gap.** Although promising, the application of MDE to AI in healthcare (MDE4AI) is still in its early stages. Existing approaches usually tackle isolated problems but rarely deliver end-to-end solutions that are interoperable and privacy-preserving. Specifically:

- Many prototypes emphasize code generation without covering the full development lifecycle.
- Semantic interoperability is only partly addressed, with limited integration of standards such as HL7 FHIR or SNOMED CT.
- Federated learning (FL) has seldom been combined with MDE, leaving issues of privacy, governance, and reproducibility largely unsolved.

**Research Questions**. This paper addresses these shortcomings through three guiding questions:

- RQ1: How can MDE principles be applied to design healthcare AI platforms that ensure interoperability, privacy, and traceability?
- RQ2: Can a healthcare-focused DSL allow clinicians to define machine learning pipelines without extensive programming expertise?
- RQ3: How effective is a model-driven, federated approach in producing accurate predictive models across multiple clinical sites?

**Contributions.** To answer these questions, we propose a model-driven framework for AI-enabled healthcare, implemented in the Smart Digital Health Platform (SDHP). Our contributions are:

1. The Medical Interoperability Language (MILA): a DSL that incorporates clinical ontologies and HL7 FHIR resources to guarantee semantic consistency across heterogeneous datasets.
2. Integration with Federated Learning: a model-driven pipeline that ensures consistent pre-processing and training at all sites, enabling privacy-preserving multi-center analytics.
3. Empirical Evaluation: a cancer immunotherapy case study across four European clinical centers, showing that the framework achieves strong predictive performance (support vector machines up to 99.35% and 81.69% on key tasks), reduces manual coding effort, and provides end-to-end traceability.

The remainder of this paper is organized as follows. Section 2 reviews related work on MDE, semantic interoperability, and federated learning in healthcare AI. Section 3 introduces the proposed MDE4AI framework, outlining its principles and four main stages (Model Definition, Validation, Transformation, and Code Generation & Deployment). Section 4 details the implementation of the framework, including the MILA toolchain, ontology integration, and federated learning setup. Section 5 presents the QUALITOP case study, describing the experimental setup, model performance across multiple prediction tasks, and validation of traceability and development effort. Section 6 discusses the benefits, clinical and regulatory relevance, limitations, and future opportunities of MDE4AI. Finally, Section 7 concludes by summarizing key contributions and outlining directions for extending the framework to multimodal and explainable healthcare AI.

## 2 RELATED WORK

T The integration of AI into healthcare faces challenges in interoperability, data privacy, and system complexity, which have been addressed in prior work through three key approaches: (1) Model-Driven Engineering (MDE) and Domain-Specific Languages (DSLs), which streamline development but lack mature end-to-end frame-works for AI (discussed in Section 2.1); (2) Semantic Interoperability, where standards like FHIR help reconcile heterogeneous data but struggle with inconsistent implementations (discussed in Section 2.2); and (3) Federated Learning (FL), which enables privacy-preserving collaboration but remains hindered by ad hoc designs and governance gaps (discussed in Section 2.3). This section synthesizes these themes, identifying the need for cohesive MDE frameworks that unify DSLs, semantic harmonization, and FL for scalable healthcare AI.

### 2.1 Model-Driven Engineering and Domain-Specific Languages in Healthcare

Model-Driven Engineering (MDE) elevates healthcare software development by using high-level models, such as UML diagrams or platform-independent models (PIMs), and automated transformations to generate platform-specific models (PSMs) or code. This approach reduces manual coding effort, increases traceability, and bridges the gap between clinicians and technical implementations. A core advantage of MDE is the ability to create domain-specific languages (DSLs) tailored to healthcare, enabling clinicians to model data flows, clinical workflows, and AI analytics without deep programming knowledge. In our Smart Digital Health Platform (SDHP).

Despite its potential, **recent literature highlights the infancy of MDE4AI in healthcare**. Rädler *et al.* (2025) reviewed 1,335 articles and found only 18 primary studies integrating MDE and AI development, with DSL-based support for preprocessing and deployment still **limited in maturity** [5]. Naveed *et al.* (2024) report that many MDE-for-ML efforts focus on model syntax and code generation but **neglect data preparation, lifecycle integration, and tooling robustness**, resulting in solutions that remain largely **prototype-level** [6]. Similarly, Brandon *et al.* (2023) presented exploratory case studies of MDE in AI-enabled healthcare, such as diagnostics and patient monitoring, but noted the **absence of comprehensive, end-to-end frameworks** combining intuitive DSLs, metamodeling, and automated AI pipeline generation. Key adoption barriers persist, including **complex DSLs, rigid or cumbersome tooling, and poor integration with clinical workflows**, all of which can undermine participatory development by healthcare professionals [7].

**Scientific gap:** There is a **lack of usable, lifecycle-complete MDE frameworks for healthcare AI**, especially solutions that combine DSLs, semantic modeling, and automated code generation to deliver **clinically accessible, end-to-end pipelines**.

This paper contributes by a **MDE4AI framework** with MILA as a healthcare-specific DSL. MILA's metamodel embeds standards like **FHIR**, ensuring semantic consistency, and the MDE pipeline supports **code generation, data integration, and deployment**. By providing a **simplified DSL syntax and user-friendly tools** (e.g., graphical editors and automated generators), our approach aims to overcome previous adoption and usability challenges, enabling **intuitive, end-to-end model-driven development for healthcare AI**.

### 2.2 Semantic Interoperability in Healthcare

The ability of different health systems to exchange data with consistent meaning, remains a persistent challenge in smart health platforms. Without a shared semantic framework, data from one hospital's EMR may be misinterpreted by another system, especially in AI-driven or cross-border settings. Standards like HL7 FHIR (Fast Healthcare Interoperability Resources) have been developed to address this issue by providing structured resource models (e.g., Patient, Observation, Medication) and standardized terminology. The adoption of such standards is critical to ensuring that patient data remains meaningful and usable across diverse EHR systems. International initiatives, such as the European Health Data Space and WHO digital health strategies, underscore the centrality of semantic interoperability to health data integration [8], [9].

Despite the availability of standards like FHIR and SNOMED CT, achieving full semantic interoperability is far from solved. A 2024 scoping review noted persistent challenges in FHIR adoption, including terminology mapping issues, inconsistent implementations, and limited scalability. Systematic reviews report that real-world platforms often require significant manual mapping and data transformation, hindering usability by domain experts. Bossenko et al. (2024) demonstrate tools that require significant manual mapping and data transformation, limiting real-world usage by domain experts despite promising tooling for EHR integration. Highlighting the complexity of FHIR resources and the absence of governance models as key obstacles [10]. Rigas et al. (2024) emphasize the need for multi-level approaches combining standardized models, terminologies, and curated datasets, calling for more real-world studies to evaluate practical interoperability [11].

**Scientific gap:** many existing systems stop short of full semantic integration, offering limited support for end-to-end automation and requiring substantial effort to normalize data across institutions.

This paper contributes by embedding FHIR resources and healthcare ontologies directly into the metamodel of a model-driven smart digital health platform (SDHP). This allows modelers such as clinicians or data engineers to define queries, data flows, and AI tasks in MILA (our domain-specific language) using semantically aligned data types. Once the initial alignment to the ontology is done, the platform automatically generates semantically consistent artifacts (e.g., data schemas, API definitions) that adhere to FHIR and other standards. This approach supports semantic interoperability by design, ensuring that applications generated by the platform share a consistent data dictionary and maintain traceability of standards across the model transformation pipeline. It reduces the need for manual mapping and enhances scalability, moving toward truly interoperable healthcare systems.

### 2.3 Federated Learning

Federated learning (FL) has emerged as a promising technique for training AI models on distributed healthcare data without centralizing sensitive patient information. Unlike traditional machine learning, which requires pooling data on a central server, FL keeps data local: each participating hospital trains a model on its own data and only shares model updates (e.g., gradients) with a central aggregator. This design helps mitigate legal and ethical issues around patient privacy, one of the most pressing barriers to multi-site AI training, especially under regulations like HIPAA and GDPR. A 2024 systematic review confirms that data privacy is the top concern in collaborative health AI, and FL directly addresses this by design [12], [13], [14], [15].

Despite this potential, real-world adoption of FL in healthcare remains limited. A 2024 analysis of 612 FL studies found that only 5.2% had reached clinical deployment, with most being proof-of-concept prototypes [16]. A follow-up review by Saeed et al. (2025) points to challenges like data skew, fairness, and quality control, all of which reduce model robustness [17]. FL systems also face technical hurdles such as data heterogeneity (different formats, labels, or class balances across sites), security threats (e.g., adversarial updates or model inversion attacks), and difficulties integrating FL into existing hospital IT infrastructures. Furthermore, ethical and governance concerns persist even with privacy-preserving design, institutions must agree on how models are trained, validated, and used, who owns the model, how bias is handled, and who is accountable for outcomes.

**Scientific gap:** most FL efforts are **ad hoc, hand-coded, and hard to reproduce**, lacking support for managing heterogeneity, enforcing governance, and ensuring transparency.

This paper contributes by addressing this gap through a **model-driven engineering (MDE)** approach. We explicitly model the federated architecture in the platform's **Platform-Independent**

**Model (PIM)**, capturing each institution (or node) with its data types, resources, privacy policies, and governance constraints. The central coordinator is also modeled as a distinct component with defined roles. Data heterogeneity is handled upfront by mapping all node schemas to a shared ontology (as covered in the semantic interoperability section), and from these models, we automatically generate harmonized pipelines, adapters, and preprocessing logic that ensure model compatibility across sites.

### 2.4 Condensed Gap-Analysis Table (systems level)

Table 1 Comparative gap analysis of prior work

| *Category of Prior Work* | *Representative Studies* | *Observed gap* | *Our bridge* |
|---|---|---|---|
| *MDE/DSLs* | Declarative extraction for AI on FHIR/CDMs [18]<br>Clinical NLP service [19]<br>Docker/OMOP/FHIR training & deployment [20] | Clinician-usable end-to-end modeling still missing; DSLs/tools remain engineer-centric. | MILA + MDE pipeline: healthcare-specific DSL, full lifecycle (prep→train→deploy→monitor). |
| *Semantic Interop* | Smart-hospital FHIR platform [21]<br>FHIR Mapping Language tooling) [22]<br>FHIR analytics framework [23] | Semantics handled at data layer, not embedded into modeling/automation. | Semantics-by-design: FHIR/ontologies in the metamodel/DSL; generators emit FHIR-consistent artifacts. |
| *Federated Learning* | Federated data/analytics [24]<br>multi-center FL with FHIR sync [25]<br>FL on FAIR/FHIR [26] | Ad-hoc data harmonization; limited governance; no DSL for clinician-authored FL tasks. | Model-driven FL: PIM-level federation, auto-generated adapters, governance hooks, clinician DSL. |

Contribution of this paper: In contrast to systems that address only one dimension, MDE/DSLs, semantics, or FL, our framework unifies all three as presented in Table 1: a healthcare-specific DSL (MILA) expresses tasks with FHIR-typed semantics; the MDE pipeline compiles them into executable code and APIs; and federated orchestration is generated from platform-independent models with built-in governance and auditability.

## 3 METHODS AND FRAMEWORK

This section presents the model-driven engineering (MDE) framework that underpins our approach to AI-powered healthcare platforms. The framework is designed to bridge the gap between high-level clinical requirements and executable machine learning pipelines by combining three key principles: **abstraction through models, semantic alignment via ontologies, and privacy-preserving collaboration enabled by federated learning**. Following established practices in model-driven engineering, the framework proceeds in four main stages, **Model Definition, Model Validation, Model Transformation, and Code Generation & Deployment**, which progressively refine abstract specifications into operational pipelines. At the center of this process is the Medical Interoperability Language (MILA), a domain-specific language that captures analytical intent in a structured, semantically grounded form. Each stage of the framework engages MILA alongside the Clinical Ontology, the Virtual Data Lake, and the Federated Learning layer, ensuring that the resulting pipelines remain interoperable, reproducible, and faithful to clinical objectives. The following subsections describe each stage in turn.

### 3.1 Overview of the MDE4AI Framework

Figure 1 illustrates the proposed model-driven engineering (MDE) pipeline for healthcare AI, which is organized into four main stages: **Model Definition, Model Validation, Model Transformation, and Code Generation & Deployment**. These stages represent the progressive refinement of high-level clinical specifications into an executable and deployable AI pipeline.

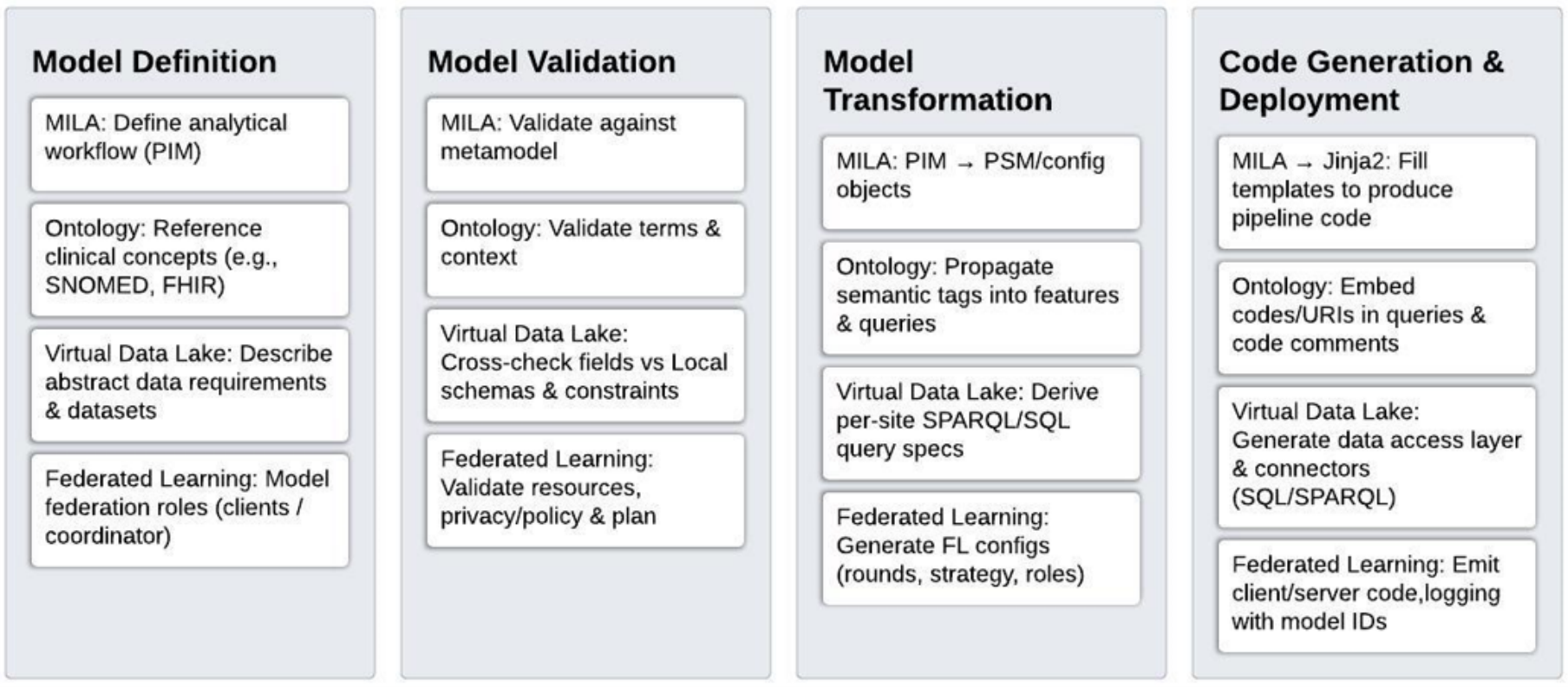

Figure 1 The model-driven engineering (MDE) pipeline for healthcare AI.

At the heart of the framework is the **MILA domain-specific language (DSL)**, which allows experts to describe analytic workflows at a high level of abstraction. MILA is complemented by three additional components: the **Clinical Ontology**, which provides standardized semantics for all data elements; the **Virtual Data Lake**, which abstracts access to heterogeneous datasets across sites; and the **Federated Learning layer**, which supports privacy-preserving distributed training. Each stage of the pipeline engages these components in different ways to ensure that the resulting system is faithful to the clinical intent, semantically consistent, and operationally feasible.

- In the **Model Definition** stage, clinicians and data scientists describe analytical goals and data needs in MILA, referencing ontology-backed clinical concepts while remaining agnostic to storage and infrastructure.
- During **Model Validation**, the model is checked for correctness, semantic integrity, and data availability across the federation, with safeguards for privacy and resource constraints.

- The **Model Transformation** stage then converts the abstract specification into platform-specific designs and configurations, resolving data access through the Virtual Data Lake and generating federated learning parameters.
- Finally, **Code Generation & Deployment** produces executable scripts and connectors, embedding ontology references for traceability and enabling distributed execution across participating sites.

This stepwise progression ensures that an abstract clinical model can be systematically translated into an operational AI pipeline without losing semantic meaning or violating privacy requirements. Each of the following subsections (3.2–3.5) expands on these stages in detail, highlighting how MILA, the ontology, the Virtual Data Lake, and federated learning interact at each point in the pipeline.

### 3.2 The MILA DSL and Metamodel Design

The first step in the pipeline is the **Model Definition** stage, which is expressed through the Medical Interoperability Language (MILA). MILA is a graphical domain-specific language (DSL) designed to allow clinical and data experts to describe analytical workflows in a structured but intuitive way. Unlike traditional programming, MILA raises the level of abstraction: instead of writing code, experts define the **analytical objectives, data needs, and processing steps** using visual constructs that map directly to clinical concepts.

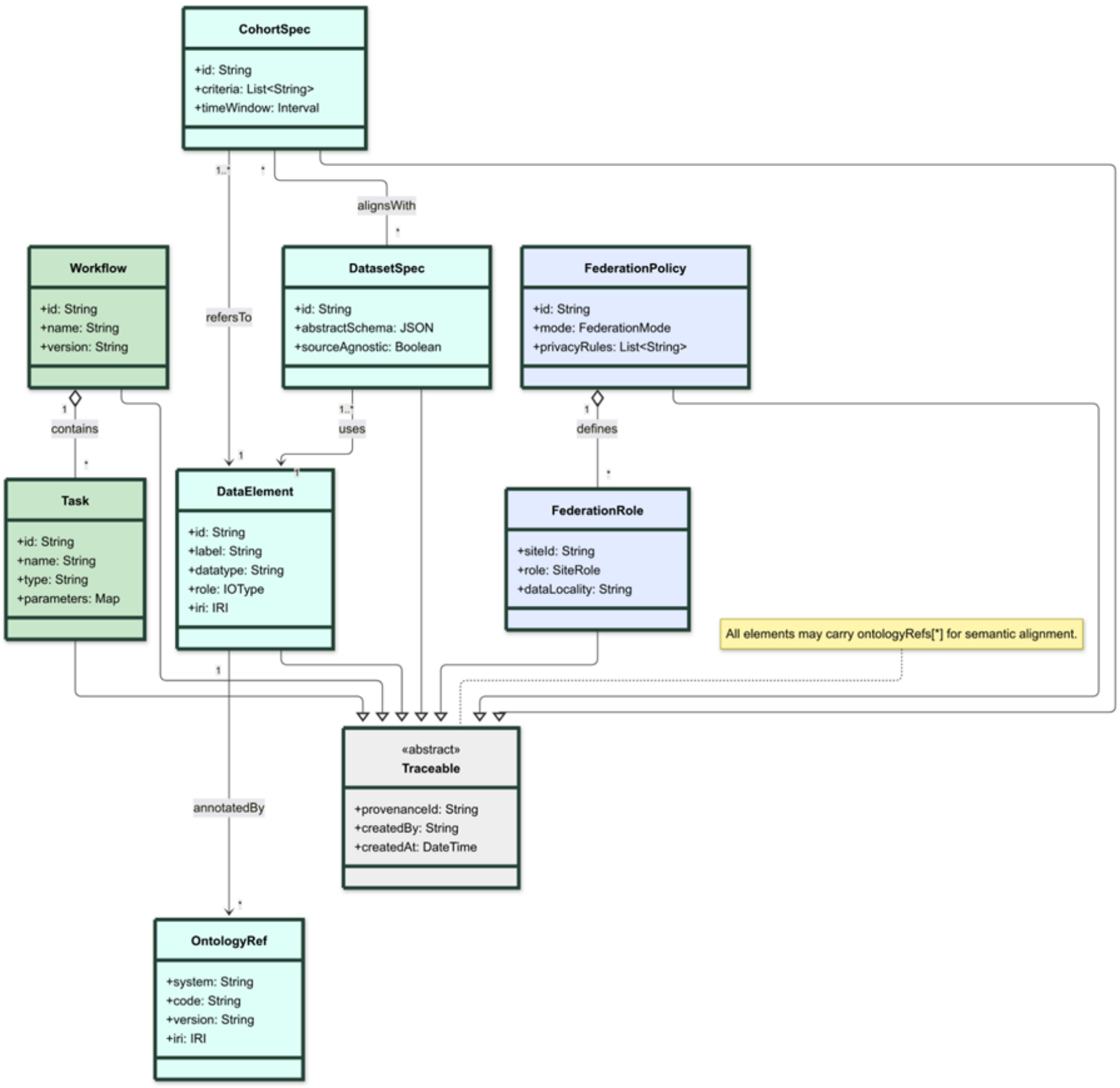

Figure 2 Meta-Model of MILA Language modelled as a class diagram

At its core, MILA is defined by a **metamodel** as presented in Figure 2, which specifies the main building blocks of the language. These include:

- **Workflows and Tasks**, which describe the steps of an analysis (e.g., preprocessing, feature extraction, model training).
- **Data Elements**, which are linked to standardized concepts in clinical ontologies such as SNOMED CT and HL7 FHIR. This ensures that inputs and outputs are described unambiguously and remain interoperable across sites.
- **Cohort and Dataset Specifications**, which allow abstract requirements to be stated (e.g., "patients with metastatic melanoma who received immunotherapy within the last 12 months"). These remain agnostic to the physical data sources, which are resolved later by the Virtual Data Lake.

- **Federation Roles**, which define whether a workflow is to be executed locally, across multiple hospitals, or in a federated setting. This makes data locality and privacy constraints explicit from the outset.

By embedding these constructs into a formal metamodel, MILA enables systematic transformations into downstream artifacts. Each model created in MILA is by definition a **platform-independent model (PIM)**: it captures *what* the analysis should do, but not *how* it is technically implemented. This separation ensures that the same MILA workflow can later be deployed in different technical environments (e.g., SQL-based systems, RDF triplestores, or federated learning setups) without rewriting the specification.

Importantly, MILA also supports **traceability and semantic alignment**. Because every data element or outcome is annotated with ontology identifiers, the meaning of variables is preserved throughout the pipeline, even after transformation into executable code. This guarantees that the pipeline faithfully reflects the clinical intent and can be audited by both technical and medical stakeholders.

Once the MILA model is specified, the next step is to validate it for correctness, semantic soundness, and feasibility across federated sites.

### 3.3 Semantic Interoperability Layer

The **Model Validation** stage is where the abstract MILA specification is checked for both correctness and feasibility before it can be transformed into an executable design. Validation operates on several levels as illustrated in Figure 3, combining structural checks with semantic reasoning and data governance rules.

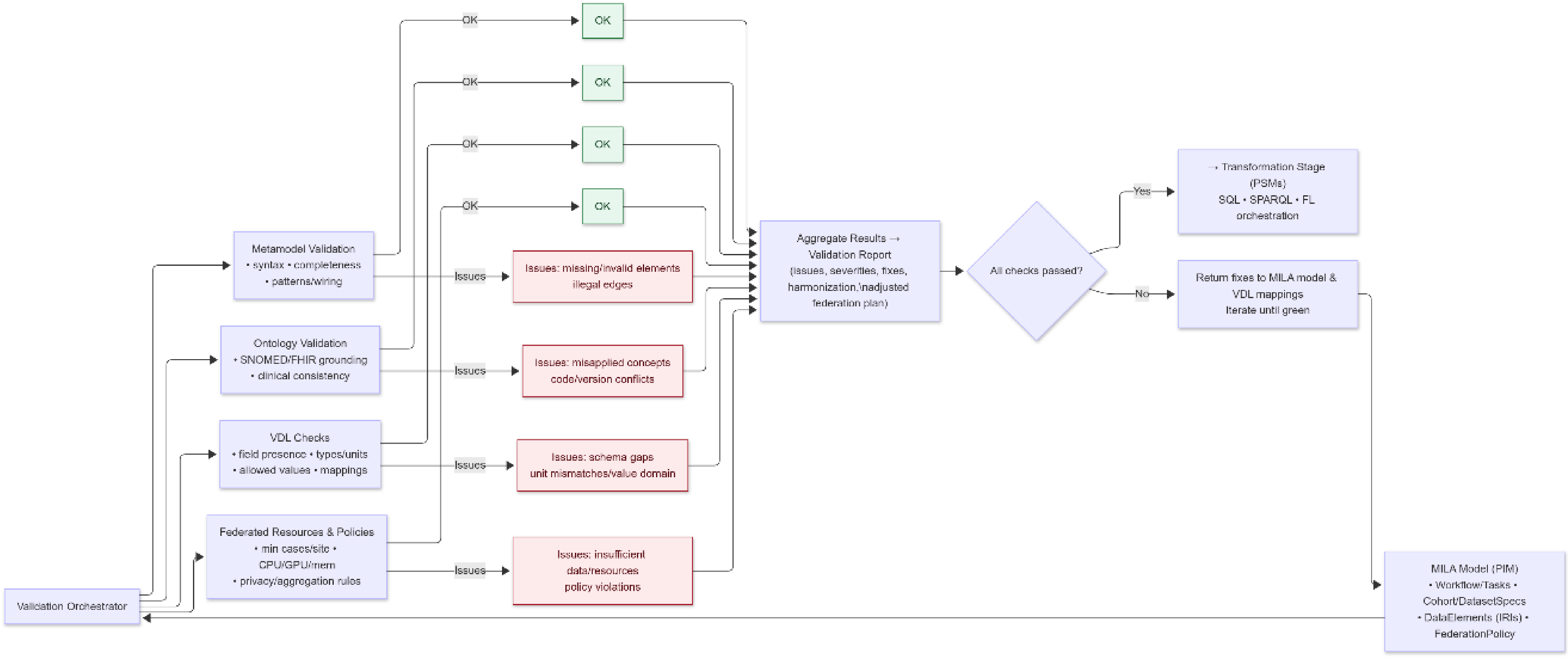

Figure 3 Validation of MILA workflow

First, the MILA workflow is validated against its **metamodel**. This ensures that the specification is syntactically correct and complete: all required elements are present, the workflow structure follows allowed patterns, and no invalid connections exist between tasks.

Second, validation extends to the **Clinical Ontology**. Each clinical concept used in the model—whether a symptom, laboratory test, or diagnosis—is checked to ensure that it is applied in an appropriate context. For example, if a workflow specifies that hypertension can be predicted from "tumor markers," the ontology would flag this as a clinical inconsistency. By grounding every concept in standardized vocabularies such as SNOMED CT and HL7 FHIR, the framework ensures that data elements are both interoperable and clinically sound. The detailed design and validation of this semantic layer are reported in our companion work on ontology-driven digital twins for chronic care [27]. I n this paper, we focus specifically on how that layer integrates with MILA, the Virtual Data Lake, and federated learning to support model-driven AI pipelines.

Third, the **Virtual Data Lake** acts as a bridge between abstract requirements and actual data schemas across participating sites. Here, each referenced data element is cross-checked against the available resources. This involves verifying not only that the field exists in local schemas, but also that its data type, unit of measurement, and allowed values match the expectations of the model. In this way, interoperability problems—such as one hospital recording blood glucose in mg/dL while another uses mmol/L—are detected and harmonized before pipeline execution.

Finally, validation considers the **federated learning context**. The system checks that each client site has sufficient data instances for local training, and that policy requirements are met. Privacy-preserving rules, such as prohibiting the transfer of identifiable patient data, are enforced at this stage by adjusting federation plans and ensuring that only aggregated model updates will be exchanged.

By combining these layers of validation, the framework guarantees that a MILA model is not only syntactically correct but also semantically meaningful, implementable on real-world data, and compliant with privacy and governance constraints. Only models that pass these checks advance to the transformation stage, reducing the risk of failure or inconsistency later in the pipeline.

Only models that pass these validation steps move forward into the transformation stage, where they are prepared for execution in a federated setting.

### 3.4 Federated Learning Architecture

After the validation stage, the platform-independent model (PIM) defined in MILA is systematically transformed into a **platform-specific model (PSM)** that can be executed across multiple sites. This step is not simply a technical translation, but a conceptual bridge: it ensures that the abstract clinical requirements expressed in MILA are faithfully mapped to concrete workflows, data access mechanisms, and learning protocols that can run in a distributed environment.

As demonstrated in **Error! Reference source not found.** , each component of the framework plays a distinct role. The **MILA DSL** defines the analytical workflow in a structured form, while the **Clinical Ontology** ensures that every feature and outcome remains semantically anchored during transformation. For example, if a model includes "hypertension" as an input feature, the ontology identifier is carried forward into the design so that the meaning of the feature is preserved across sites and into the executable code.

The **Virtual Data Lake** resolves abstract data requirements into site-specific retrieval logic. Rather than requiring hospitals to harmonize their physical databases in advance, the system automatically generates tailored queries for each site (e.g., SQL for relational databases or SPARQL for RDF triplestores). This approach maintains semantic uniformity while respecting local heterogeneity, allowing institutions to collaborate without changing their internal data infrastructure.

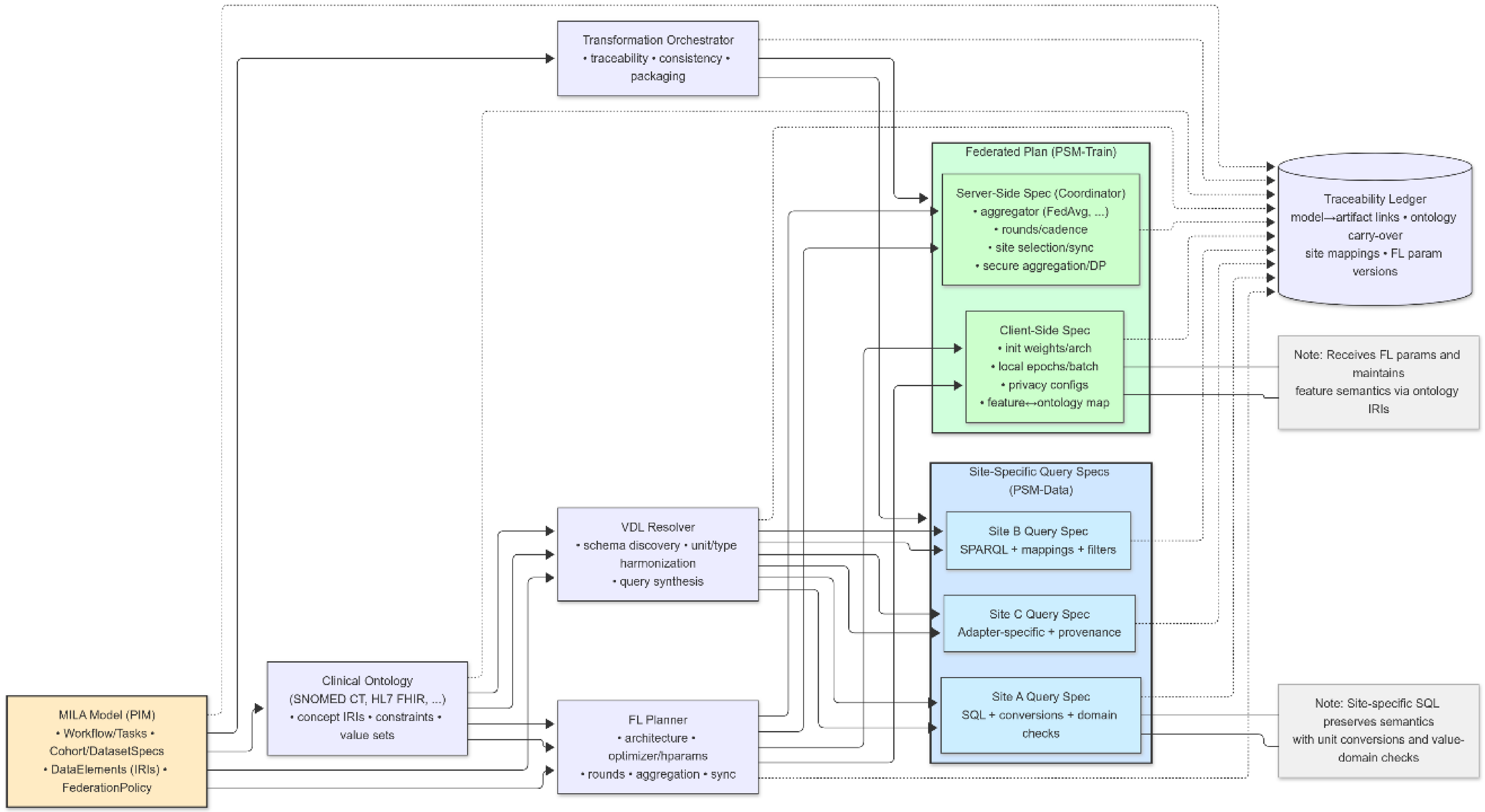

Figure 4 Federated Learning Architecture

In parallel, the **Federated Learning (FL) layer** is configured. The transformation process generates the specifications that govern distributed training, including the choice of aggregation algorithm (e.g., Federated Averaging), the number of training rounds, and synchronization protocols between the coordinator and client sites. By encoding these parameters directly from the high-level MILA model, the framework guarantees that all participating sites use identical model architectures and training configurations. This avoids the inconsistencies and manual errors that often undermine multi-center studies.

Conceptually, this stage embeds **traceability and uniformity** into the federation. Every data element can be traced back to its ontology reference, every query is derived from the same abstract specification, and every local model is trained under the same global protocol. The outcome of the transformation is thus more than a set of configuration files: it is a complete, federated-ready design that faithfully implements the clinical and analytical intent expressed at the modeling level.

### 3.5 MDE Pipeline: From Definition to Deployment

Figure 5 shows the final stage of the pipeline is **Code Generation and Deployment**, where the platform-specific model (PSM) is transformed into a fully executable AI pipeline. This stage brings the abstract design to life: what began as a high-level MILA specification is now realized as concrete code, directly runnable within each participating institution.

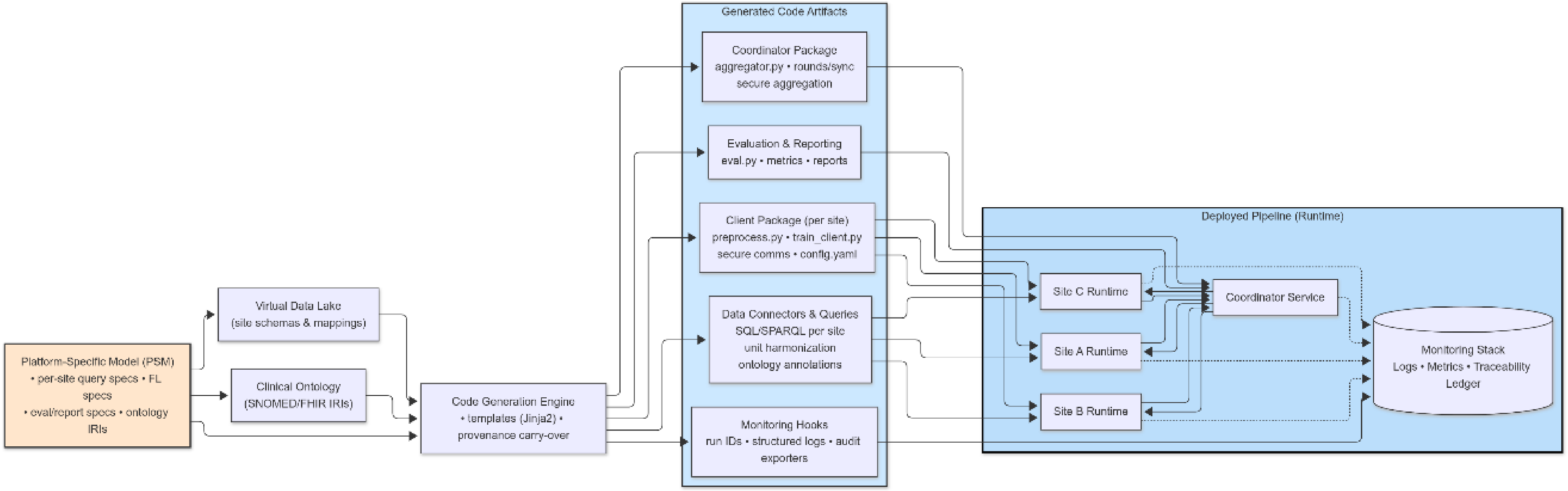

Figure 5 MDE Pipeline: From Definition to Deployment

Using a **template-based approach** (e.g., Jinja2), the framework automatically generates source code modules for data preprocessing, model training, evaluation, and reporting. Each piece of code is annotated with ontology identifiers, ensuring that provenance remains transparent. For instance, when a feature such as *blood pressure* is extracted from a dataset, the generated code explicitly carries the SNOMED CT or HL7 FHIR identifier, so that downstream outputs can always be traced back to their clinical definition. This practice strengthens auditability and provides confidence that the system adheres to established standards.

The **Virtual Data Lake (VDL)** contributes by generating the data-access layer. Instead of hard-coded connections, the system produces connectors and queries tailored to each institution's database schema. This enables heterogeneous systems to participate in a uniform pipeline, abstracting away local technical differences while maintaining semantic consistency.

For **federated learning setups**, the code generator produces a distributed suite of scripts rather than a single monolithic program. Each hospital receives a **client-side package** responsible for local preprocessing, training, and secure communication, while the central coordinator runs an **aggregation routine** (e.g., Federated Averaging) to combine model updates. This separation guarantees that patient-level data never leave the local environment, while still enabling collaborative training across sites.

To support **traceability and monitoring**, the generated code incorporates experiment identifiers, structured logs, and audit hooks. Each training round can be linked back to the MILA specification that produced it, enabling regulators, developers, or clinicians to reconstruct the decision path of the system.

Finally, the generated artifacts are executed **directly within hospital infrastructures**. There is no need for additional packaging or orchestration layers in the current framework; instead, the code can be run as-is in the respective runtime environments. The result is an operational AI pipeline that respects data privacy, runs seamlessly across heterogeneous environments, and preserves the semantic integrity of the original MILA model.

In this way, the MDE4AI framework closes the gap between clinical abstraction and technical execution, ensuring that every deployed pipeline remains faithful to its original intent.

## 4 IMPLEMENTATION

This section describes the technical implementation of the proposed MDE4AI framework, showing how the conceptual principles outlined in Section 3 are realized in practice. The implementation is built around a **model-driven toolchain** that transforms high-level MILA specifications into executable Python pipelines, ensures **semantic consistency** through ontology integration, and enables **privacy-preserving training** via federated learning. The design emphasizes modularity, automation, and traceability: changes made at the model level are automatically propagated to code, semantic annotations are carried through to runtime, and federated services are deployed in a uniform way across sites. The following subsections present the toolchain and model specification process (Section 4.1), explain how semantic interoperability is enforced in practice (Section 4.2), and describe the federated learning deployment across multiple clinical centers (Section 4.3).

### 4.1 Toolchain and Model Specification

The implementation of MILA follows a model-driven development workflow, where abstract specifications are progressively transformed into executable code. As described conceptually in Section 3, MILA models capture analytic workflows at the **platform-independent level (PIM)**. In practice, these workflows are represented as **JSON files** that conform to a metamodel defined in the **Eclipse Modeling Framework (EMF)**. This ensures that each model instance is structurally valid and aligns with the formal syntax of the DSL.

Once defined, these JSON models are automatically converted into **platform-specific implementations (PSMs)** through a model-to-text transformation. We implemented this process using **Python generation scripts** and **Jinja2 templates**, which render parameterized code skeletons into runnable modules. The generated artifacts include preprocessing routines, training logic, and evaluation scripts, all tailored to the specified workflow. This approach not only automates code production but also guarantees consistency: when a high-level MILA model is updated, the pipeline code can be regenerated seamlessly, keeping all downstream artifacts in sync.

The structure of the implementation is illustrated in Figure 6, which shows how model specifications, templates, and runtime code are cleanly separated. The app/models/ directory contains JSON definitions of workflows (e.g., *Treatment_prediction.json*, *Adverse_Event_causation.json*). Corresponding templates are stored under templates/, where each workflow type is defined as a parameterized code skeleton (e.g., *Treatment_prediction_pipeline.j2*). Running the generation script instantiates these templates with model-specific details, producing Python pipeline code. The generated services are placed under app/services/, while REST API routes are defined in app/routes/, exposing the workflows as callable endpoints.

This modular structure improves **maintainability and traceability**. For example, ontology identifiers and model IDs are embedded directly into the generated code as comments or log entries, enabling any runtime operation to be traced back to the original model element. Such design practices, common in safety-critical domains, are particularly important in healthcare AI, where reproducibility and auditability are essential. In summary, the MILA toolchain combines EMF-based validation, JSON specifications, and Jinja2-driven code generation to ensure that every abstract model can be translated into consistent, deployable Python pipelines with a transparent audit trail.

```
MILA_MODELS/
├── app/
│   ├── models/
│   │   └── Treatment_prediction.json
│   │   └── Adverse_Event_caustion.json
│   │   └── Treatment_causse_AE.json
│   │   └── AE_prediction.json
│   ├── routes/
│   │   └── __init__.py
│   │   └── treatment.py
│   │   └── AdverseEvents.py
│   ├── services/
│   │   └── base_service.py
│   │   └── treatment_service.py
│   │   └── AdverseEvents_service.py
│   └── ...
├── templates/
│   └── Treatment_prediction_pipeline.j2
│   └── Adverse_Event_caustion_pipeline.j2
│   └── Treatment_causse_AE_pipeline.j2
│   └── AE_prediction_pipeline.j2
├── run.py
```

Figure 6 Directory structure of the MILA_MODELS project.

### 4.2 Semantic Interoperability in Practice

Semantic interoperability is a cornerstone of the MILA framework, ensuring that data and analytic tasks retain consistent meaning across different institutions and technical environments. As outlined in Section 3.3, the semantic layer builds on standardized clinical ontologies such as SNOMED CT and HL7 FHIR, and its design and validation are reported in detail in our companion work on ontology-driven digital twins for chronic care (Elgammal et al., under review in *Software and Systems Modeling*). In the current paper, we focus on its practical role within the implementation.

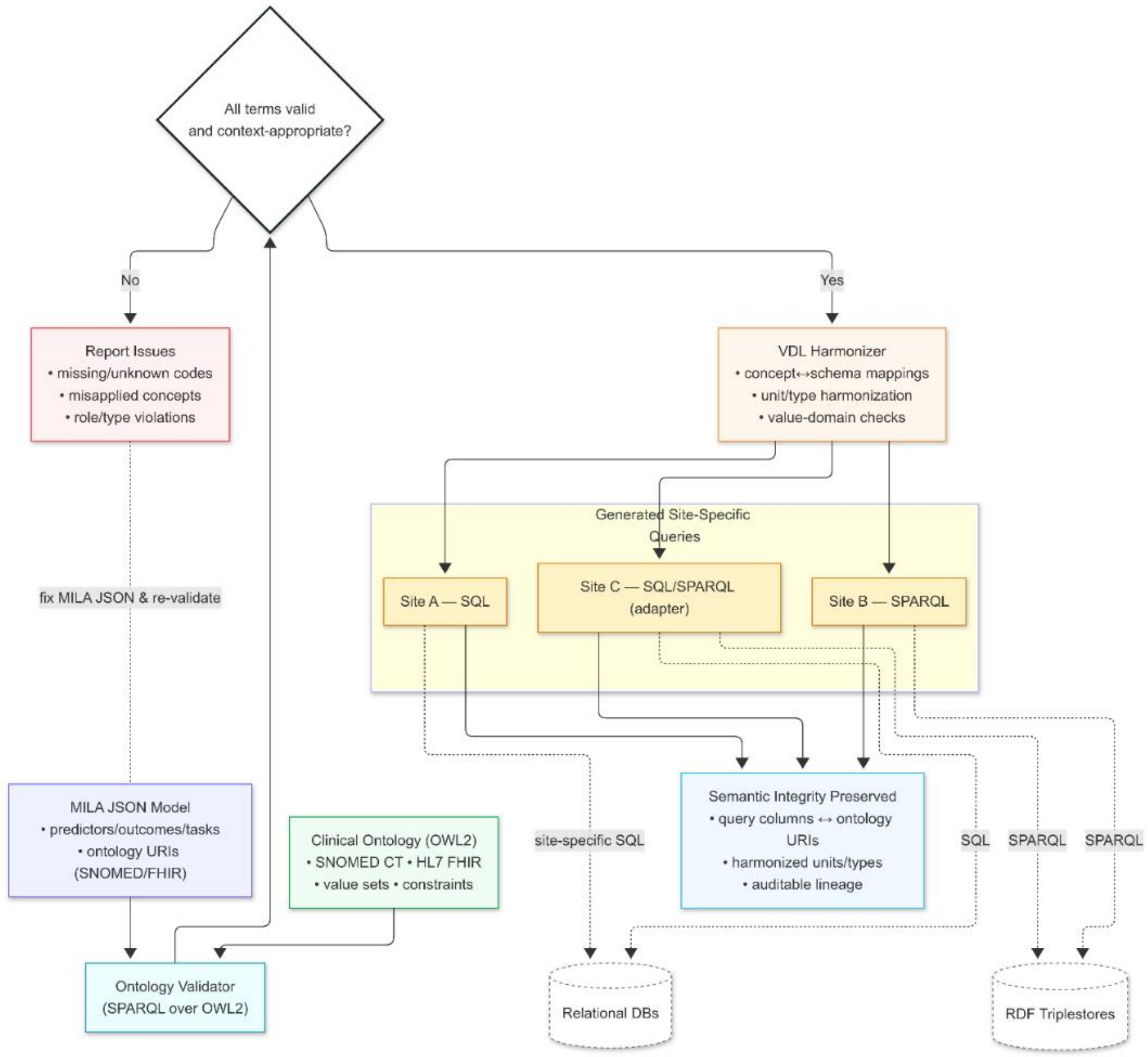

Figure 7 Semantic Interoperability in Practice

As illustrated in Figure 7, each MILA JSON model embeds **ontology references** directly into its definitions of predictors, outcomes, and tasks. For example, a workflow predicting treatment response may specify “blood pressure” or “immune checkpoint inhibitor therapy” not as free-text labels, but as URIs linked to the appropriate ontology concepts. These annotations are validated against an OWL2-based ontology using SPARQL queries, ensuring that every term exists and is applied in the correct context. This prevents inconsistencies such as assigning a laboratory test to an outcome role or mislabeling a symptom as a treatment.

Beyond harmonizing terminology, semantic interoperability also extends to **data access**. The MILA model describes data needs abstractly, using ontology-backed variables rather than database-specific fields. At deployment time, these abstract requirements are resolved into **site-specific queries**: SQL for relational systems or SPARQL for RDF triplestores. For instance, if a model specifies “adverse event incidence” as a feature, the generator will produce different queries for each hospital depending on its database schema, while still guaranteeing semantic equivalence. This approach resembles ontology-based data access, where mappings connect high-level specifications to heterogeneous sources without requiring institutions to modify their internal data structures.

A key benefit of this design is that **semantic integrity is preserved throughout the pipeline**. Whether data is pulled from a relational table or a graph database, it is always linked back to a standard ontology concept, making the resulting models interpretable and auditable across sites. In federated deployments, this ensures that all participating hospitals are training on semantically consistent features, even if their underlying databases are vastly different.

### 4.3 Federated Learning Deployment

To enable privacy-preserving model training across institutions, the MILA framework adopts a **Federated Learning (FL) architecture**. Building on the conceptual design described in Section 3.4, the implementation follows the classical Federated Averaging (FedAvg) strategy, which allows models to be trained collaboratively without centralizing data. As presented in Figure 8.

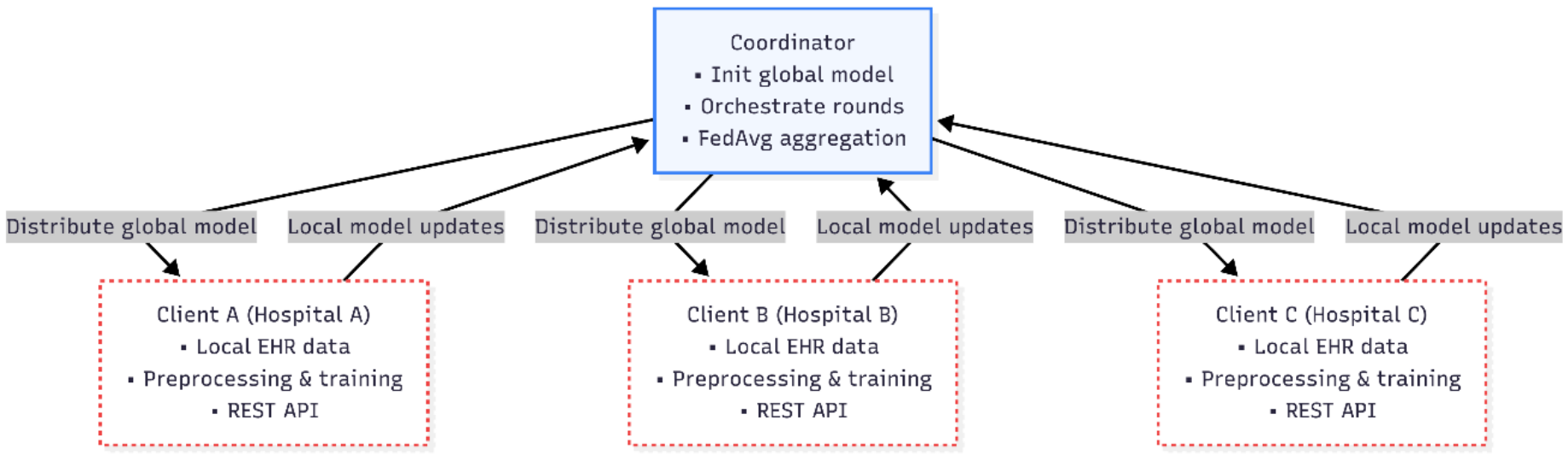

Figure 8 Federated Learning Deployment

In practice, the generated pipeline code is deployed at each participating hospital within its secure environment. Each site runs the MILA-derived client package, which includes preprocessing, local training, and lightweight communication routines. At the start of a session, an initial global model is distributed to all sites. Each hospital then performs local training epochs on its own electronic health record (EHR) data using the generated preprocessing and training code. Instead of transmitting raw data, each site returns only model updates (weights or gradients).

These updates are aggregated using the FedAvg algorithm to produce a new global model that incorporates knowledge from all sites. The updated model is redistributed for the next training round. Because all sites share the same MILA-derived pipeline code, the training process is guaranteed to be uniform — every local update is computed under the same assumptions, hyperparameters, and workflow logic. This uniformity removes a common source of error in federated learning, where manually coded pipelines may diverge in subtle ways.

From an implementation perspective, the framework cleanly separates site-specific configuration (e.g., database connectors, local compute settings) from the federated workflow logic. Each hospital operates its client service within its own environment, while the aggregation logic can be executed in a trusted process external to the clients. To support traceability, every training update and evaluation metric is tagged with model identifiers and site IDs, making it possible to audit contributions and verify the provenance of the final aggregated model.
This automated pipeline was tested across multiple clinical centers, where the system successfully carried out synchronous training rounds. The results demonstrate the practical viability of the framework: a single high-level MILA specification can be transformed into distributed, privacy-preserving learning code that operates consistently across heterogeneous healthcare settings.

## 5 CASE STUDY: QUALITOP EVALUATION AND RESULTS

To evaluate the effectiveness of the proposed MDE4AI framework, we conducted a case study using real-world data from the **EU H2020 QUALITOP project**, a multi-country cohort study on cancer immunotherapy. This setting provided a challenging testbed: data were distributed across four clinical sites, spanned multiple modalities, and were subject to strict privacy requirements. The case study allowed us to examine not only whether the framework could achieve high predictive accuracy, but also whether it could deliver on its core promises of **semantic interoperability, federated reproducibility, traceability, and reduced development effort**. This section presents the experimental setup (Section 5.1), evaluates model performance (Section 5.2), validates uniformity and traceability (Section 5.3), and compares development effort with traditional coding approaches (Section 5.4).

### 5.1 Experimental setup

We evaluated the MDE4AI framework in the context of the EU H2020 **QUALITOP** project, which conducted an international, observational multicenter cohort study in **France, the Netherlands, Portugal, and Spain** [28]. The study focused on patients receiving **CAR-T cell therapy** or **immune checkpoint inhibitors**, two immunotherapies that are clinically complex due to the frequent occurrence of immune-related adverse events (irAEs) and their impact on long-term quality of life (QoL). This real-world setting provided a diverse and challenging environment to test whether our framework could manage heterogeneous data while supporting federated predictive analytics.
Patients were followed for up to **18 months post-treatment**, with data collected at baseline and follow-up visits (3, 6, 12, and 18 months). The dataset included **multimodal evidence** spanning clinical records, laboratory measurements, registries, administrative data, and patient-reported QoL outcomes. Such heterogeneity reflects the real challenges of healthcare AI and makes QUALITOP a suitable benchmark for evaluating whether MDE principles — semantic modeling, platform-independent workflows, and federated execution — can produce reproducible and interoperable pipelines.

For this evaluation, we deployed the generated pipelines across **four clinical centers**. Each site retained its own patient data locally; no raw data was exchanged. Instead, the federated learning layer orchestrated distributed training, allowing local models to be trained independently and aggregated centrally using Federated Averaging. To stress-test the framework, we defined **five predictive tasks**, covering treatment outcomes and adverse event monitoring. These tasks required combining clinical and longitudinal data, thereby exercising MILA's semantic layer, the Virtual Data Lake's schema mappings, and the federated execution pipeline.
This setup allowed us to evaluate the framework along three axes:

- **Predictive performance** — could standardized federated pipelines achieve competitive accuracy across sites?
- **Uniformity and traceability** — were the generated workflows consistent across hospitals, and could model outputs be traced back to their MILA specifications?
- **Development effort** — how much manual coding was avoided by using MILA compared to a conventional implementation?

### 5.2 Model performance

We evaluated four oncology-related prediction tasks (Models A–D) on the federated QUALITOP dataset, using a uniform pipeline across four European clinical sites. Each task addressed a different question: Model A – treatment recommendation, Model B – whether an adverse event (AE) was treatment-related, Model C – detection of treatment-induced AEs, and Model D – prediction of a patient's future AE family. All pipelines were generated from the same MILA specifications, ensuring consistent preprocessing and feature definitions across sites. For brevity, we focus on accuracy as the primary metric in the main text presented in Table 2; other performance metrics (F1, weighted F1, ROC-AUC) followed similar patterns and are illustrated in Figure 9.

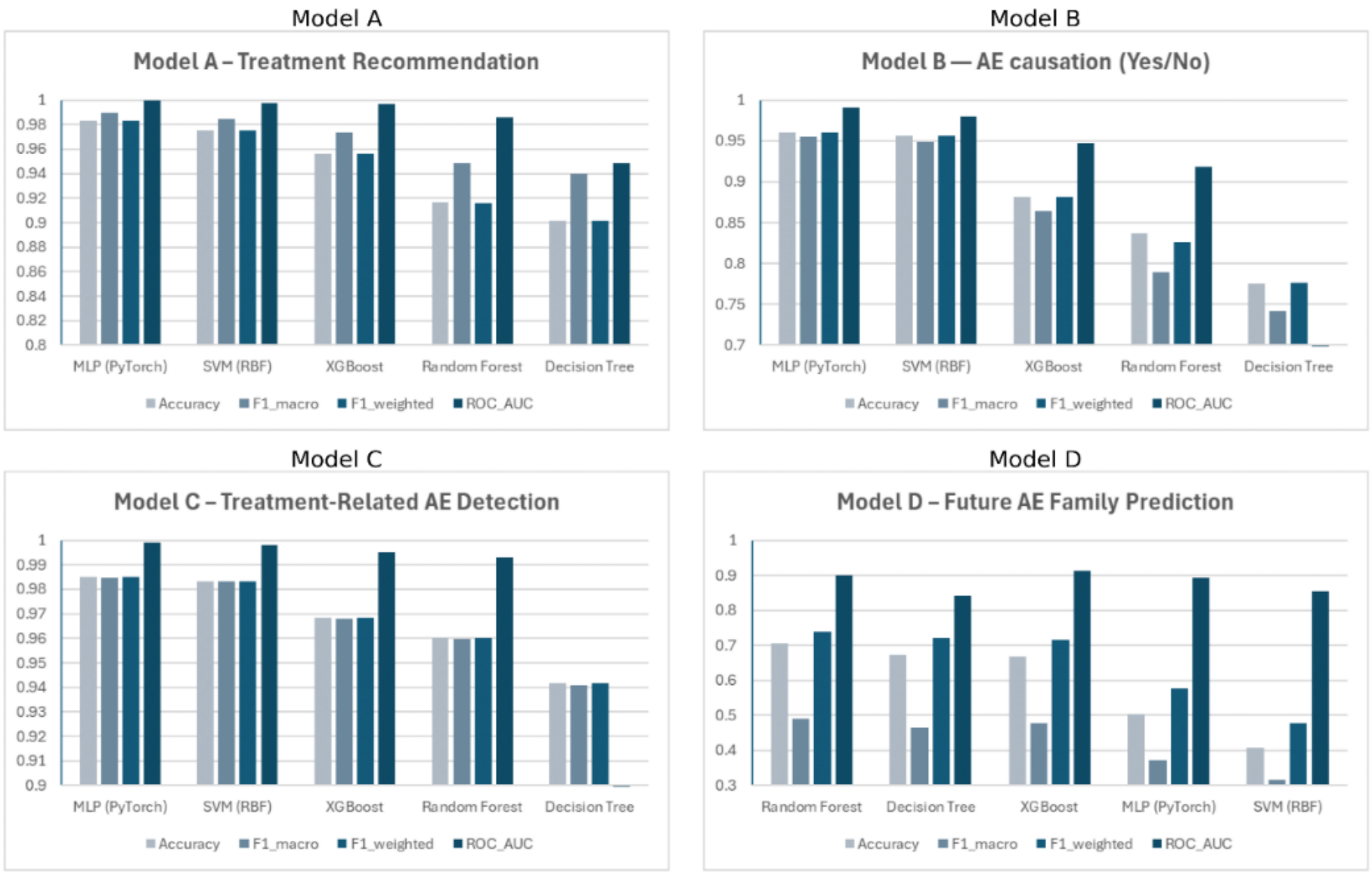

Figure 9 Performance comparison of four predictive models across different clinical tasks. Each subplot shows evaluation metrics (Accuracy, F1-macro, F1-weighted, and ROC-AUC) for multiple algorithms. (A) Model A – Treatment recommendation. (B) Model B – Adverse

On Models A, B, and C, the Multi-Layer Perceptron (MLP) and RBF-kernel SVM consistently achieved the highest accuracy (approximately 95–98%). These two models also yielded near-perfect F1-scores and ROC-AUC values (≈0.99) in these tasks, indicating excellent all-around performance. The XGBoost classifier performed slightly lower (accuracy in the high 80s to mid-90s), while the ensemble tree-based methods (Random Forest and Decision Tree) lagged further behind. For instance, in treatment recommendation (Model A), MLP achieved 97.7% accuracy versus 91.7% for Random Forest. A similar trend appeared in AE causality (Model B) and treatment-related AE detection (Model C), where MLP and SVM reached around 96–98% accuracy, substantially outperforming Random Forest (83–96%) and Decision Tree (77–94%) on those tasks. These results indicate that the more flexible models (neural network and SVM) can effectively capture the structured treatment–outcome relationships inherent in tasks A–C.

In contrast, task D (future AE family prediction) proved much more challenging, and here the performance rankings inverted. The highest accuracy was achieved by the Random Forest (about 70.7%), outperforming both Decision Tree (~67.3%) and XGBoost (~66.8%). The MLP and SVM, which excelled in earlier tasks, struggled on task D – achieving only 48.8% and 40.7% accuracy, respectively. This suggests that for complex longitudinal predictions like future AE families, simpler ensemble methods generalized better than more expressive models. All accuracy results are summarized in Table 2 Accuracy comparison across models and tasksbelow. Notably, adopting a model-driven, uniform pipeline did not hinder predictive power. On the contrary, our framework's accuracies are comparable to or even higher than those reported in similar manually coded studies. This demonstrates that MDE4AI can serve as a robust backbone for federated healthcare AI applications, delivering high performance while maintaining consistency across sites and tasks.

Table 2 Accuracy comparison across models and tasks

| MODEL | MODEL A (TREATMENT) | MODEL B (AE CAUSALITY) | MODEL C (TREATMENT-AE) | MODEL D (FUTURE AE FAMILY) |
|---|---|---|---|---|
| **MLP (PYTORCH)** | 0.977 | 0.961 | 0.985 | 0.488 |
| **SVM (RBF)** | 0.975 | 0.956 | 0.983 | 0.407 |
| **XGBOOST** | 0.957 | 0.881 | 0.968 | 0.668 |
| **RANDOM FOREST** | 0.917 | 0.837 | 0.960 | 0.707 |
| **DECISION TREE** | 0.902 | 0.776 | 0.942 | 0.673 |

Importantly, the results demonstrate that model-driven uniformity did not constrain predictive power. On the contrary, the framework achieved performance comparable to or better than manually coded studies in similar domains, indicating that MDE4AI can serve as a reliable backbone for federated healthcare AI applications.

### 5.3 Traceability and uniformity validation

A defining strength of the proposed MDE4AI framework is its ability to guarantee **uniformity and traceability** across distributed deployments. Because all pipelines were generated from a single MILA specification, each clinical site executed the same preprocessing, feature selection, and training steps. This removed one of the main pitfalls in federated learning, where local deviations often cause model divergence. In our experiments, the FedAvg aggregation converged smoothly, confirming that the consistent model structure simplified federation and reduced error sources.

Traceability was validated through an audit of the four prediction tasks (Models A–D). For each task, we confirmed that predictions could be traced backward through the generated Python code into the corresponding JSON model, and ultimately to the original MILA specification with preserved ontology references. This ensured that all outputs remained explainable in terms of their underlying design. The audit results are summarized in Table 3, where all tasks passed full traceability checks.

Table 3 Results of the traceability audit across selected prediction tasks.

| PREDICTION TASK (MODEL) | TRACED TO JSON MODEL? | TRACED TO MILA SPECIFICATION? | ONTOLOGY REFERENCE PRESERVED? |
|---|---|---|---|
| MODEL A – TREATMENT RECOMMENDATION | ✓ | ✓ | ✓ |
| MODEL B – AE CAUSALITY | ✓ | ✓ | ✓ |
| MODEL C – TREATMENT-RELATED AE | ✓ | ✓ | ✓ |
| MODEL D – FUTURE AE FAMILY | ✓ | ✓ | ✓ |

Each prediction could be traced back through the generated code to the JSON model and the original MILA specification, with ontology references preserved. This confirms that the MDE4AI framework provides end-to-end transparency and reproducibility.

### 5.4 Development effort comparison

In addition to accuracy and traceability, we assessed the **development effort required** to implement new predictive tasks with and without the MDE4AI framework. Traditionally, creating a new machine learning pipeline in healthcare involves manual steps such as data parsing, feature engineering, model configuration, and deployment scripting. This process can take several days of coding and is error-prone, especially when the pipeline must be replicated consistently across multiple institutions.

Table 4 Development effort required for new predictive tasks: manual coding vs. MDE4AI pipeline.

| DEVELOPMENT ACTIVITY | MANUAL CODING EFFORT | MDE4AI EFFORT (MILA + GENERATION) |
|---|---|---|
| DATA PARSING & CONNECTORS | Days of custom scripting | Automated from Virtual Data Lake |
| FEATURE ENGINEERING | Manual coding per dataset | Declared in MILA specification |
| MODEL CONFIGURATION | Hand-written hyperparameters | Defined once in MILA model |
| CODE IMPLEMENTATION | 100s–1000s lines of code | Auto-generated from templates |
| TOTAL TIME (PER TASK) | Days to weeks | Hours (JSON authoring only) |

With MDE4AI, these tasks were greatly simplified. Instead of writing code manually, developers only needed to author a new **MILA model specification** in JSON. From this high-level description, the system automatically generated the preprocessing routines, model code, and service interfaces. As a result, the time and effort required to implement new tasks were reduced by an order of magnitude, while ensuring that semantic annotations and traceability links were preserved.

The comparison in Table 4 highlights this difference. While manual coding required substantial effort across all stages, the MDE-driven approach reduced the workload to model authoring and automated generation, improving both **efficiency and reproducibility**.

This evaluation demonstrates that MDE4AI not only improves **accuracy and reproducibility**, but also substantially reduces the **cost and time of development**, a factor that is critical for scaling AI solutions in clinical practice.

### 5.5 Summary of Results

Taken together, the QUALITOP case study demonstrates that the MDE4AI framework can deliver not only **competitive predictive performance** across federated datasets, but also **uniformity, traceability, and reduced development effort**. By ensuring that all sites executed identical, ontology-anchored workflows, the framework avoided inconsistencies that often undermine multi-center AI studies. The traceability audit confirmed that every prediction could be linked back to its original model specification, satisfying key requirements for transparency and clinical auditability. Finally, the effort comparison showed that high-level modeling in MILA dramatically reduced the time and complexity of implementing new tasks compared to manual coding. These results provide a strong indication that MDE-driven approaches can make federated healthcare AI both **technically robust** and **practically scalable**.

## 6 DISCUSSION

The evaluation of the MDE4AI framework in the QUALITOP case study demonstrated both strong technical performance and practical feasibility. Beyond reporting predictive accuracy, it is important to reflect on what these results mean for the broader challenges of deploying AI in healthcare. In this section, we discuss the benefits of the framework (semantic alignment, reproducibility, and traceability), its clinical and regulatory relevance, the limitations that currently constrain its generalizability, and the future opportunities that could extend its impact. This structured discussion helps position the work within the larger context of healthcare informatics and highlights directions for continued development.

### 6.1 Framework Benefits

The evaluation of MDE4AI in the QUALITOP case study highlights several benefits of adopting a model-driven approach for healthcare AI. First, the framework achieved **semantic alignment** across heterogeneous datasets by embedding clinical ontologies such as SNOMED CT and HL7 FHIR directly into the modeling process. This ensured that concepts such as diagnoses, laboratory results, and outcomes were represented consistently across sites, reducing ambiguity and enabling reliable integration of multimodal evidence.

Second, the use of MILA and automated code generation enforced **reproducibility in federated learning**. As presented in Section 5.3, every site executed identical workflows derived from a single high-level specification, eliminating discrepancies in preprocessing, feature engineering, or model structure. This uniformity simplified the aggregation process in FedAvg and avoided the model divergence that often affects multi-center AI studies.

Third, the framework provided **end-to-end traceability**, allowing predictions to be linked back through the generated code into the MILA specification and associated ontology references. The traceability audit Table 3 confirmed that this lineage was preserved for multiple tasks, offering a level of accountability rarely available in hand-coded pipelines.

Finally, MDE4AI delivered measurable gains in **development efficiency**. As presented in Section 5.4, adding a new predictive task required only the creation of a MILA model rather than days of manual coding. This reduced the time to deployment while maintaining transparency and semantic consistency, making the framework practical for scaling AI across diverse clinical studies.

Together, these benefits suggest that MDE4AI provides a foundation for healthcare AI systems that are not only technically robust but also aligned with the clinical and regulatory demands of transparency, interoperability, and reproducibility.

### 6.2 Clinical and regulatory relevance

Beyond technical performance, the MDE4AI framework addresses requirements that are central to the **clinical and regulatory acceptance** of AI systems. One of the most pressing challenges in healthcare AI is ensuring compliance with **privacy regulations** such as the General Data Protection Regulation (GDPR) in Europe and the Health Insurance Portability and Accountability Act (HIPAA) in the United States. By adopting a federated learning architecture, our framework avoids the need to centralize patient-level data, ensuring that sensitive information remains within each institution's secure environment. This design directly supports data protection mandates while still enabling collaborative model development.

Equally important is the demand for **auditability and transparency** in clinical AI. Healthcare providers, regulators, and patients increasingly expect that AI-driven decisions can be explained and verified. MDE4AI inherently supports this through its traceability features: every model decision can be linked back to its original MILA specification and associated ontology references (as presented in Section 5.3). This level of accountability is aligned with recent regulatory guidance that emphasizes explainability as a prerequisite for trustworthy AI in medicine.

Finally, the framework supports **interoperability standards** such as HL7 FHIR and SNOMED CT, which are increasingly required for electronic health record (EHR) integration and digital health certification. By embedding these standards into the modeling process, MDE4AI reduces the risk of vendor lock-in and increases the likelihood that the system can be integrated into existing clinical infrastructures without extensive customization.

Taken together, these features demonstrate that MDE4AI is not only a technical framework but also a **regulatory- and practice-aware approach**. Its alignment with privacy, auditability, and interoperability requirements increases its relevance for healthcare organizations seeking to deploy AI responsibly and at scale.

### 6.3 Limitations and Challenges

While the case study results are encouraging, several limitations of the current MDE4AI framework should be acknowledged. First, the evaluation was conducted in the context of **oncology immunotherapy** using the QUALITOP dataset. Although this setting provided rich and heterogeneous data, the framework's generalizability to other clinical domains has yet to be demonstrated. Broader testing in cardiology, primary care, or multimorbidity contexts will be needed to confirm its versatility.

Second, the current implementation primarily addresses **structured clinical data** such as laboratory values, diagnoses, and patient-reported outcomes. Many high-value healthcare datasets, including medical images, genomic profiles, and free-text clinical notes, remain outside the scope of MILA's current metamodel. Extending the framework to handle unstructured modalities will require additional modeling constructs and integration with natural language processing (NLP) and imaging pipelines.

Third, while federated learning offers strong privacy guarantees, its effectiveness depends on the **quality and representativeness of local data**. Differences in data completeness or coding practices across sites may still affect model performance, even if workflows are uniform. Addressing these issues may require additional mechanisms for harmonization and quality control.

Finally, building the initial **metamodel and ontology mappings** requires substantial upfront effort from both technical and clinical experts. While this investment pays off in long-term automation and traceability, it may limit adoption in smaller institutions without the necessary expertise or resources.

These limitations do not undermine the promise of MDE4AI but instead highlight important areas where further development and validation are needed to ensure broad clinical applicability.

### 6.4 Future Directions

Looking ahead, several opportunities could extend the scope and impact of the MDE4AI framework. One direction is the integration of **large language models (LLMs)** into the modeling pipeline. LLMs could assist clinicians in authoring MILA specifications through natural language prompts or automatically generate candidate workflows based on textual clinical guidelines. Such integration would lower the barrier to entry and further align model-driven engineering with clinical practice.

Another priority is enhancing **explainability and interpretability**. While the current framework ensures traceability, additional tools are needed to communicate model decisions to clinicians in an intuitive way. Embedding model-agnostic explainability methods, such as SHAP or counterfactual explanations, into the generated pipelines could help bridge the gap between formal auditability and everyday clinical decision support.

The framework could also be expanded to cover a wider range of **clinical tasks and data modalities**. This includes extending MILA to support imaging, genomics, and free-text clinical notes, as well as incorporating multimodal learning techniques. Such developments would increase the framework's utility in diverse clinical domains and strengthen its position as a general-purpose infrastructure for healthcare AI.

Finally, further research should explore the integration of MDE4AI with **emerging regulatory and governance frameworks** for trustworthy AI. As clinical AI becomes more tightly regulated, systems that can demonstrate compliance by design—through standardized models, traceability, and automated audit trails—will have a strong advantage in real-world adoption.

Taken together, these opportunities suggest that MDE4AI can evolve from a promising prototype into a **next-generation platform for trustworthy digital health**, supporting not only technical robustness but also clinical usability, explainability, and regulatory readiness.

In summary, the discussion highlights that the MDE4AI framework delivers tangible benefits in terms of **semantic alignment, federated reproducibility, traceability, and development efficiency**, while also aligning closely with **clinical and regulatory requirements**. At the same time, its current scope remains limited to structured oncology data, and further work is needed to extend its coverage, validate generalizability, and lower the entry barrier for adoption. Future opportunities, particularly the integration of **LLMs, explainability tools, and multimodal data support**, offer promising pathways to strengthen both the technical capabilities and clinical utility of the framework. Together, these reflections suggest that MDE4AI provides a strong foundation for advancing trustworthy healthcare AI, while also laying out a roadmap for ongoing refinement and broader adoption.

## 7 CONCLUSION

In this paper, we presented a model-driven engineering (MDE) framework for AI-powered healthcare platforms and demonstrated its feasibility through a multi-center oncology case study. At the center of the framework is the **MILA domain-specific language (DSL)**, which enables clinical experts to specify analytical workflows at a high level of abstraction while embedding standardized ontologies for semantic consistency. Supported by the **Virtual Data Lake** and a **federated learning layer**, MILA produces pipelines that are reproducible across institutions, privacy-preserving by design, and traceable from clinical intent to executable code.

The QUALITOP case study provided **initial evidence** that the approach is both technically robust and efficient. The generated pipelines achieved competitive predictive performance (with support vector machines reaching up to 99.35% accuracy), while maintaining uniformity across distributed sites and reducing development effort to the authoring of MILA models rather than extensive manual coding. These results suggest that MDE4AI can help overcome long-standing barriers in healthcare AI, including data heterogeneity, reproducibility across sites, and regulatory demands for auditability.

At the same time, the current framework remains limited to structured oncology data and requires substantial upfront modeling effort. Future work will need to extend MILA to support unstructured modalities, integrate explainability mechanisms, and explore synergies with emerging approaches such as large language models. By addressing these challenges, MDE-driven methods could provide a foundation for **next-generation digital health platforms** that are transparent, interoperable, and clinically trustworthy.